\documentstyle[twoside,fleqn,espcrc2,psfig]{article}
\title{%
Generalized ensemble algorithm for U(1) gauge theory
}

\author{Tetsuya Takaishi\address{Hiroshima University of Economics,
 Hiroshima, 731-0192, JAPAN}
}

\def\be{\begin{equation}}
\def\ee{\end{equation}}
\def\bea{\begin{eqnarray}}
\def\eea{\end{eqnarray}}
\begin{document}
\begin{abstract}
Hybrid Monte Carlo simulations of the pure compact U(1) gauge theory 
are performed with the Tsallis weight.
The simulations show that the use of the Tsallis weight 
enhances the tunneling rate between metastable states.
%and more reliable estimate of 
%the order of the phase transition may be possible.
\end{abstract}

\maketitle

\section{INTRODUCTION}

The four dimensional pure compact U(1) gauge theory is 
known to posses a phase transition (PT)
separating a confined phase and a Coulomb one and
the determination of the order of PT is of great importance.
In the pure compact U(1) gauge theory, however,
the determination of the order of PT 
is  very difficult 
since on large lattices the standard updating algorithms like Metropolis, 
heat-bath, hybrid Monte Carlo (HMC)\cite{HMC}, 
fail to generate enough tunneling between metastable states.  
Many efforts have been done to clarify the order of the PT\cite{QED}. 
In spite of such efforts the order of the PT is still controversial.

A promising algorithm to overcome this difficulty is the multicanonical algorithm\cite{MULTI} 
which uses a multicanonical weight in stead of the Boltzmann one and can enhance 
the tunneling rate between metastable states. 
Although the multicanonical algorithm works effectively, 
one disadvantage of the multicanonical algorithm might be that
a multicanonical weight used in the simulations is not known a priori and it must be
determined before the simulations.
Usually the weight is estimated from a short run of the standard updating algorithm.
However this estimation is more difficult as a lattice size becomes larger.
 
Recently in several fields where the usual Monte Carlo technique does not work effectively, simulations were performed
with the Tsallis weight\cite{TSHMC,MC}, first introduced by Tsallis\cite{TSALLIS}.
The Tsallis weight controlled by a  parameter $q$ can be easily defined, and the usual Boltzmann
weight is given by taking the limit $q\rightarrow 1$. 
The results have showed that the use of the Tsallis weight might be more advantageous than that of 
the Boltzmann one.  

Inspired by such studies, here we apply the Tsallis weight for
the hybrid Monte Carlo simulation of 
the pure compact U(1) lattice gauge theory
and investigate whether the Tsallis weight solves the difficulty 
in determining the order of the PT.
  
\section{GENERALIZED ENSEMBLE AND TSALLIS WEIGHT}
The partition function of a system with the action $S$ is given by 
\be
Z=\int \exp(-\beta S[U])dU.
\label{Z}
\ee
Introducing a function $G$, the partition function can be rewritten as
\bea
Z & = & \int \exp(-\beta S[U])/G \cdot G dU \nonumber \\
& = & <\frac{\exp(-\beta S)}{G}>_G Z_G,
\eea
where $Z_G$ is a generalized partition function:
\be
Z_G=  \int G dU,
\ee
and $<O>_G$ stands for the expectation value of an observable $O$ in the ensemble generated 
with $Z_G$. 
The expectation value of $O$ in the canonical ensemble generated with eq.(\ref{Z}) 
can be obtained in terms of the generalized ensemble  as
\bea
<O> & = & \int O\exp(-\beta S[U])dU/Z \nonumber \\ 
    & = & <O\frac{e^{-\beta S}}{G}>_G/<\frac{e^{-\beta S}}{G}>_G.
\eea

Previously it was suggested that in some cases Tsallis weight works better than the Boltzmann one\cite{TSHMC,MC}.
In the Tsallis formulation, 
\be 
G=[1-(1-q)\beta \bar{S}]^{\frac{1}{(1-q)}},
\ee
or in the familiar exponential form,
\be
G=\exp(-\beta S_T) \equiv  W_T,
\ee
with 
\be
S_T=\frac{1}{\beta(q-1)}\ln [1-(1-q)\beta \bar{S}],
\ee
where $\bar{S}=S-S_0$. $S_0$ is a certain constant shifting the origin of the action.
Here we call $W_T$ "Tsallis weight".
The Boltzmann weight can be obtained by taking the limit  $q\rightarrow 1$ 
in the Tsallis weight $W_T$.

\section{Hybrid Monte Carlo with Tsallis weight}
The action of pure compact U(1) lattice gauge theory is given by 
\begin{equation}
S=\sum_{x,\mu>\nu} [1-\cos(\theta_{\mu\nu}(x))],
\label{ACTION}
\end{equation}
where $\theta_{\mu\nu}(x)$ is the sum of link angles contributing 
a plaquette at a site $x$ on the $\mu\nu$ plane. 
The Hamiltonian with the Tsallis weight, 
used for HMC simulations, is defined by
\begin{equation}
H_T=\sum\frac12 p^2_i +\beta S_T.
\label{HT}
\end{equation}
Here note that $S_T$ is not always  well-defined.
Namely $1-(1-q)\beta \bar{S}$ can be negative.
However, provided that $q$ is close to one,
$1-(1-q)\beta \bar{S}$ is made positive.  

\section{RESULTS}

As an exploratory study we tested the above algorithm varying $q$ 
on an $8^4$ lattice in the vicinity of the PT ( at $\beta=1.00737$ ).
We performed HMC simulations\cite{HMC} with eq.(\ref{HT}).
$S_0$ ( = 0.37 ) was taken to be the average value of the action $S$.
In order to make comparisons, we also performed HMC simulations with the Boltzmann weight.
Typically we simulated $10^6$ trajectories for each $q$.  

\begin{figure}[htb]
\vspace{-6mm}
\psfig{figure= 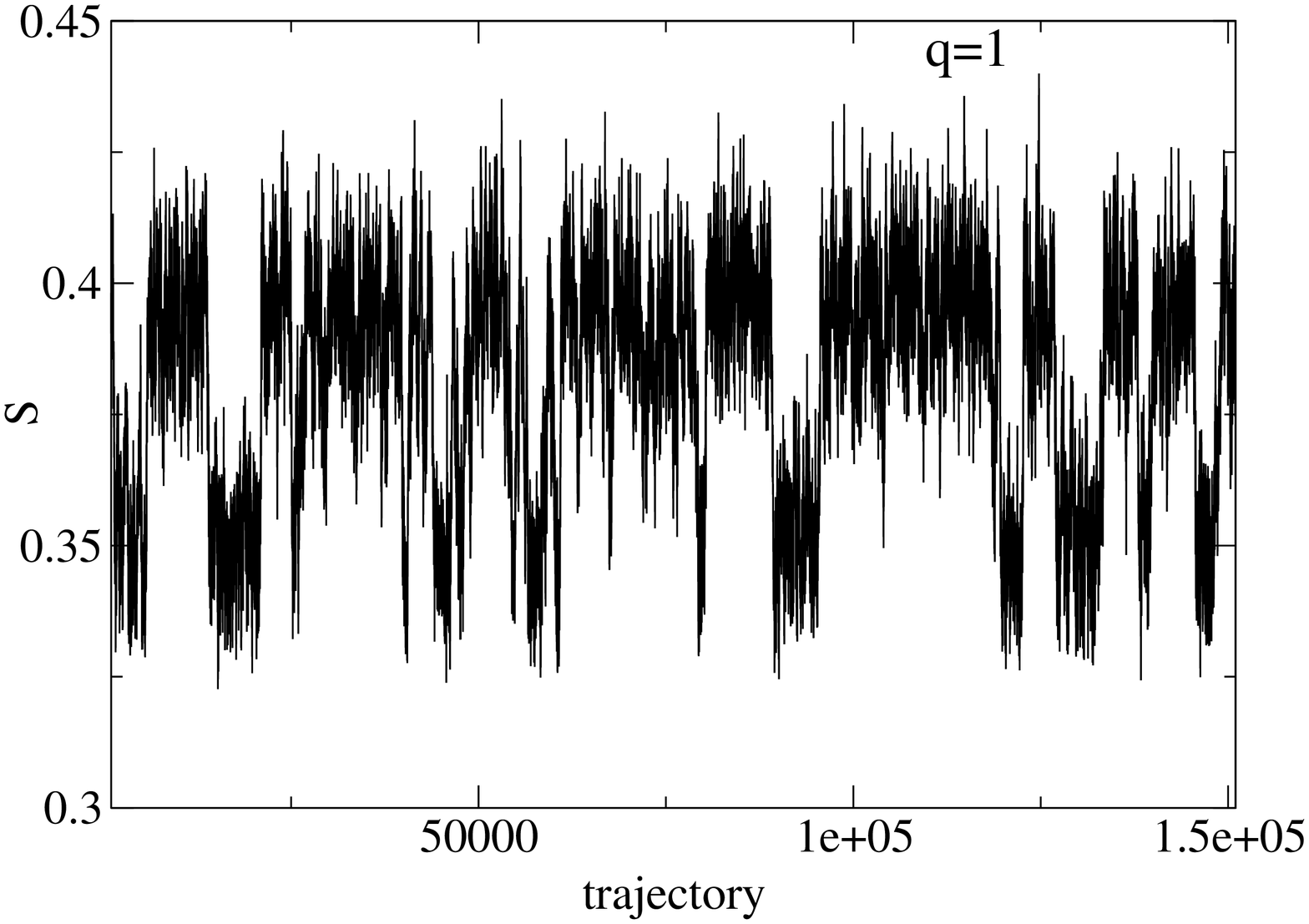,height=5.5cm,width=9cm}
\vspace{-3mm}
\psfig{figure= 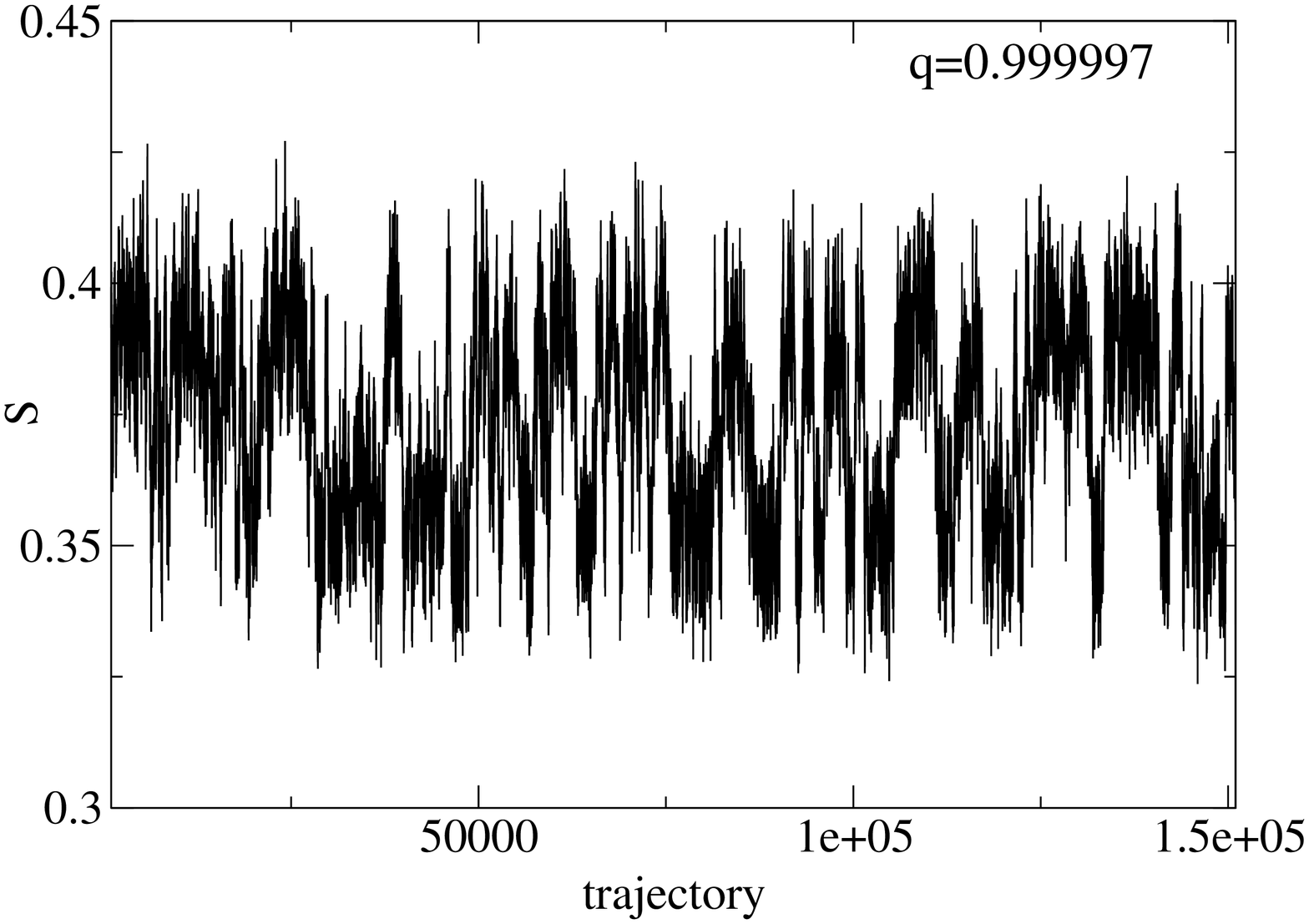,height=5.5cm,width=9cm}
\vspace{-10mm}
\caption{
(top): Typical history of action $S$ for $q=1$ ( Boltzmann weight ).
(bottom): $S_T$ for q=0.99997.
}
\vspace{-7mm}
\end{figure}

We observed more tunneling for $q<1$ than for $q=1$ (the Boltzmann weight).  
Fig.1 shows representative history of actions of q=1 and 0.999997 and
we see more tunneling for the case of q=0.999997.

Fig.2 shows action densities of q=1, 0.999997, 0.999995 and 0.99999.
The height of the valley between two peaks increases as $q$ decreases
which indicates that the tunneling rate increases as $q$ decreases. 
For too small $q$, however, values of the action
will locate near $S_0$ and in this case 
there may be a difficulty in reweighting.
Fig.3 and 4 show that the action densities reweighted to q=1 
and compare those with the Boltzmann action density ( $q=1$ ).
The reweight densities reproduce well the Boltzmann action density.
However, for $q=0.99999$ ( the smallest $q$ here ) 
the reweighted action density has large fluctuations at $S$ far from $S_0$,
which may indicate that one needs caution when using the Tsallis weight.

It is also possible to take $q>1$ in the simulations.
For $q>1$, however, the desired behavior, i.e. more tunneling,
was never  observed for our case.

\begin{figure}[ht]
\vspace{-9mm}
\psfig{figure= 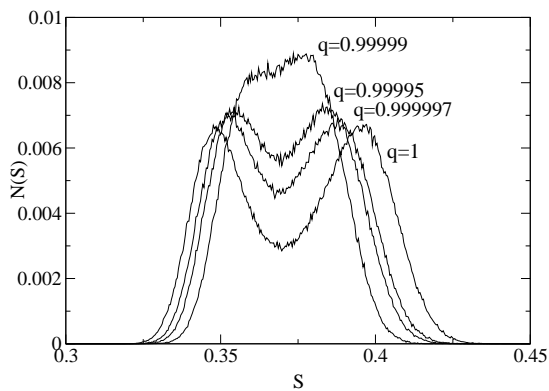,height=6.2cm}
\vspace{-12mm}
\caption{
The action densities $N(S)$ for q=1, 0.99997,0.999995 and 0.99999
}
\vspace{-10mm}
\end{figure}

\begin{figure}[ht]
\vspace{-10mm}
\psfig{figure= 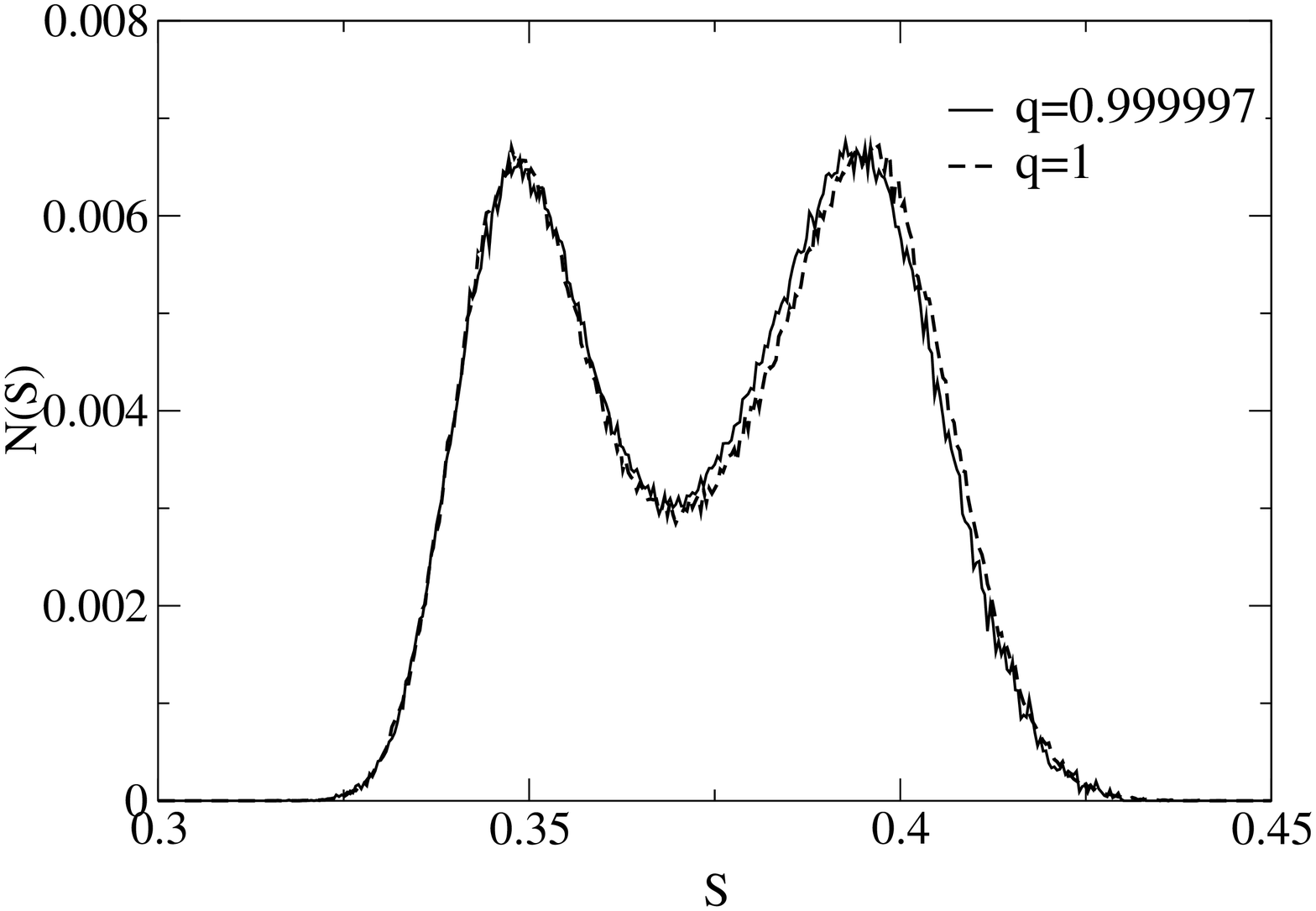,height=6.2cm}
\vspace{-12mm}
\caption{
Comparison between the action density of $q=1$ and the reweighted
action density of $q=0.999997$.
}
\vspace{-9mm}
\end{figure}

\begin{figure}[t]
\vspace{-5mm}
\psfig{figure= 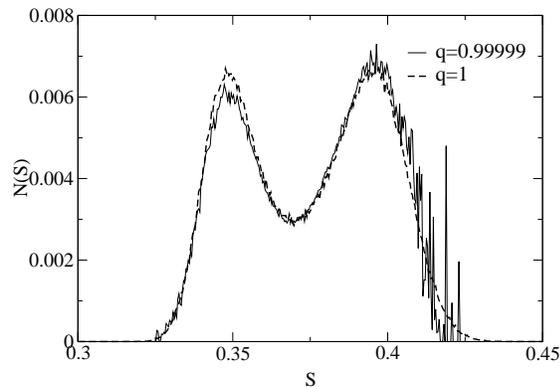,height=6.2cm}
\vspace{-12mm}
\caption{
Comparison between the action density of $q=1$ and the reweighted
action density of $q=0.99999$.
}
\vspace{-5mm}
\end{figure}

\section{DISCUSSION}
The exploratory study on an $8^4$ lattice showed that the Tsallis weight 
enhances the tunneling rate.
The lattice size used here is small.
For precise determination of the order of the PT,
larger lattices are needed.  
Therefore it is important to confirm that this preferred feature of the Tsallis weight 
is preserved for larger lattices, possibly up to $18^4$ lattice size or so. 
Simulations on larger lattices are in progress.

\section*{ACKNOWLEDGMENTS}
%\vspace{3mm}
This work was partially supported by the Ministry of Education, Science,
Sports and Culture,
Grant-in-Aid,  No.13740164.

%%%%%%%%%%%%%%%%%%%%%%%%%

\end{document}